\documentclass[twocolumn]{aa}
\usepackage{graphics}
\begin{document}
%
%   \thesaurus{08.01.1; %Stars:abundances
%              08.03.2; %Stars:chemically peculiar
%              08.09.2 3 Cen A} %Stars:individual

%
   \title{Abundances of Vanadium and Bromine in 3 Cen A\thanks{Based
on ESO spectra from the UVESPOP project}}

   \subtitle{Additional Odd-Z Anomalies}

   \author{C. R. Cowley
          \inst{1}
          \and
          G. M. Wahlgren\inst{2}
          }

   \offprints{C. R. Cowley}

   \institute{Department of Astronomy, University of Michigan,
              Ann Arbor, MI 48109-1090, USA\\
              \email{cowley@umich.edu}
         \and
             Lund Observatory, Lund University, Box 43,
             SE-22100 Lund, Sweden\\
             \email{glenn.wahlgren@astro.lu.se}
             }

   \date{Received month date, 2005; Accepted month date year}

%\maketitle

%\begin{abstract}
\abstract{We report abundance excesses of 1.2 and 2.6 dex,
respectively, for
vanadium and bromine in the hot, peculiar star 3 Cen A.
Abundances for these two odd-Z elements have not been previously
reported for this star. Taken with previous work, they strengthen the
case of the origin of the abundance peculiarities by diffusion.
 }
%\end{abstract}
  \titlerunning{Vanadium and Bromine in 3 Cen A}
  \authorrunning{Cowley and Wahlgren}

\maketitle

\section{Introduction}
%\label{se:intro}

A dwindling number of the chemical elements lack robust
identifications in stellar spectra.  Moore (1945, cf. Table 6)
omitted several elements from the Multiplet Tables as not
being -- for that time--of astrophysical interest. Among these
elements was Z = 35, bromine.  Many of the elements
listed in Moore's Table 6 have subsequently been identified
in stars.  Only recently, Castelli \& Hubrig (2004) identified
bromine in the mercury-manganese
(HgMn) star HR 7143 (HD 175640), based on three
remarkably weak features.
We note that Bidelman (2004) has suggested the
presence of Br I in the ultra-peculiar spectrum of HD 101065,
also known as Przybylski's star. In this work, we report the
identification of lines of Br II in the spectrum of 3
Centauri A (HR 5210, HD 120709). The lines are significantly
stronger than in HR 7143, though still rather weak, all less
than 6.5mA.

   We also report the unequivocal identification of V II and
abundance determination of vanadium, a spectrum identified in
many stars, but for which we have only found an upper limit
previously reported for 3 Cen A.

The findings of bromine and vanadium (Z = 23) are
of significance from the point of view
of the origin of abundance anomalies in hot, chemically
peculiar (CP) stars of the upper main sequence. The basic
mechanism postulated to account for these anomalies is a slow
diffusive
separation of the elements, depending on whether radiation
pressure pushes them upwards, or they sink under the star's
gravity. Typically, elements with high
abundances--prior to diffusion--are less
likely to be supported by radiation pressure than those with
low abundances.  This is because the lines of abundant
elements are more likely to be saturated, so the radiation
flux through them, and hence the radiation pressure on
them would be less.
This can give rise to an inverse odd-even effect
in the theoretical predictions.  Some odd-Z elements, being
typically of lower cosmic abundance than their even-Z
congeners, have larger  {\it predicted} enhancements.  The pattern
can vary for elements in different parts of the periodic table.

Perhaps
because of complicating factors, this inverse odd-even effect
is not widely observed.  However, some notorious cases can be
found in the HgMn subgroup of the CP stars.
In particular, the element gallium (Z = 31) shows huge
enhancements, while its even-Z neighbors zinc and germanium
are weak or absent.  For 3 Cen A, we can now add abundance
enhancements of the odd-Z elements vanadium and bromine.

The 3 Cen A spectrum (B5 III-IVp)
has been rather well studied (Adelman \&
Pintado 2000, Castelli, Parthasarathy, \& Hack 1997,
Wahlgren and Hubrig 2004).
The star is usually regarded as a member
of the He-weak group of CP stars. It has abundance excesses of
P (Z = 15), Sc (Z = 21), Mn (Z = 25), and especially Ga (Z = 31),
anomalies that are shared with many HgMn stars.
It is also one of the stars where the dominant helium isotope is
$^3$He (cf. Hartoog \& Cowley 1979).

Our study makes use of the high-quality spectra available from
the ESO UVES Paranal Observatory Project archive
(Bagnulo, et al. 2004, henceforth UVESPOP).
Spectra were downloaded from the web site. The
resolution is ``about 80,000."  We estimate S/N for
the unfiltered (``individual reduced")
spectra to be between 100 and 150.
%%and...typical S/N ratio is 300-500
%%in the V band.''

Two approaches to data analysis were undertaken for comparison,
both employing the Kurucz (1993) ATLAS9/WIDTH9/SYNTHE suite
of programs with updated (3 Cen A) abundances.

In one approach the merged UVESPOP ascii wavelength-flux data were
normalized, mildly Fourier filtered, and 2852 stellar features were
measured (by CRC using programs developed at the University of
Michigan).
The software records both wavelengths and the
central intensities of the features.  The measurements were
not made automatically; each required a human judgment;
the reality of the faintest measurements is often moot.
Stellar wavelengths ranged from 3070 to 10390\AA.  A subset of
the measurements, excluding lines longer than 1 micron were then
subjected to analysis by wavelength coincidence statistics
(WCS, cf. Cowley \& Hensberge 1981).  The spectra of Br II and
V II were found in this survey.  Equivalent widths were measured
by fitting individual Voigt profiles to the stellar absorption
features, and abundances obtained from WIDTH9.

The second approach relied on synthetic spectrum
modeling using two
UVESPOP ``individual reduced spectra.''  The star is
not known to be variable, and we found no indication of variability
in the spectra discussed here.
%%raw UVESPOP data.
%%GLENN: UNLESS YOU WORKED WITH THE 2D DATA, YOU WERE WORKING WITH
%%WHAT THE UVESPOP PEOPLE CALL "INDIVIDUAL REDUCED SPECTRA."
%%I'VE LOOKED AT THESE, AND THEY LOOK EXACTLY LIKE YOUR FIGURES.

The spectrum synthesis was conducted (in Lund
by GMW) using the SYNTHE program and a model atmosphere
generated using the ATLAS9 code. The model and starting set of
elemental abundances is that used by Wahlgren \& Hubrig (2004).
Included into the synthetic spectrum analysis were the hyperfine
structures for lines of both elements.

Both analyses adopted a model with $T_{eff} = 17500$K
and $\log(g) = 3.8$
as well as the overall abundances from the sources cited.

\section{Bromine}

Table 1 shows laboratory and stellar data for
Br II.  The laboratory data come from Sansonetti,
Martin, \& Young's
(2005) list of strong lines.  Intensities marked with a ``P'' are
called {\it persistent} lines.   Stellar wavelengths ($\lambda^*$)
are in column 4, given here to the fraction of
an angstrom.  A dash indicates that no wavelength was measured
near a laboratory wavelength.  Weak features may or may not be
measured, depending on rather complex subjective factors.
Classified lines, or lines whose
levels are known, are indicated by a `y'
in column 2.

%__________________________________________________ One column table

\begin{table}
\caption{Br II lines studied}
\label{tab:br2}
\begin{tabular}{ccccccc} \hline\hline
Int.& cla&$\lambda$&$\lambda^*$&$W_{\lambda}$ & log$gf$&Abund\\
        &     &     \AA        &     \AA     &     m\AA     &
&log(Ab/Sum)\\
\hline

 500 & n &3914.38&  -- &   2.3 &        &        \\
 500 & n &3980.38&  .33&   3.9 &        &        \\
1000 & n &4233.89&  -- &   $<$1.0&        &        \\
1000 & n &4365.63&  .64&   6.1 &        &        \\
 500 & n &4542.89&  -- &   $<$1  &        &        \\
 500 & n &4678.70&  -- &   $<$1  &        &        \\
 500P& y &4704.86&  .85&   6.3 &   +0.408&  $-$6.79 \\
 500P& y &4785.49&  .47&   4.7 &   +0.208&  $-$6.72 \\
 500P& y &4816.67&  .71&   2.3 &   +0.060&  $-$6.91 \\
 400 & y &4930.62&  -- &   1.3 &        &        \\
 500 & n &5182.36&  .36&   4.4 &        &        \\
 500 & y &5238.26&  .30&   $<$1  &        &        \\
 500 & y &5332.07&  -- &   1.9 &        &        \\
 500 & n &5506.72&  -- &   $<$1  &        &        \\
\hline
\end{tabular}
\end{table}

A WCS test with a tolerance window of 0.07\AA, the largest 
tolerance in the table, shows that the significance level
of seven coincidences out of the 14 wavelengths sought is 
0.00002 on the basis of 50000 Monte Carlo trials.  This 
means that in {\it just one} of the 50000 trials were 
seven or more coincidences within $\pm 0.07$ \AA\ found on 
a set of 14 nonsense wavelengths randomly placed near the 
laboratory positions. We consider the bromine 
identification to be beyond doubt from this statistic. 

Stellar absorptions near all the laboratory positions were 
examined and equivalent widths measured or, for the 
weakest features, estimated.  The most questionable 
feature is an unclassified line, $\lambda$4542.89, which 
is missing, or just short of a local maximum. 

Natural bromine consists of two isotopes found in the 
percentages of $^{79}$Br:$^{81}$Br = 50.69:49.31
(Rosman and Taylor 1998).  As both
isotopes are of odd atomic number we must be careful when 
fitting line profiles to account for
hyperfine structure (hfs). This requires both the
determination of wavelengths for individual components of 
the structure and the distribution of the total oscillator 
strength over these components. Unfortunately, relatively 
little experimental work has been conducted on the 
spectrum of Br II. The two isotopes of bromine have the 
same nuclear spin (I = 3/2) and similar magnetic moments. 
An isotope shift for Br I $\lambda$6122 has been measured to
be about 2.5 m\AA\ (Tolansky 1932), but no IS has been
observed in Br II lines (Ranade 1951). Therefore, IS is 
ignored in this analysis. 

Abundances are based on three lines for which transition
probabilities are available (Bengston \& Miller 1976).
The lines are
located at air wavelengths 4704.862, 4785.494, 4816.669
\AA\, and are transitions connecting the upper levels
4p$^3$($^4$S$^o$)5p $^5$P$_{3,2,1}$, respectively, to the 
lower level 4p$^3$($^4$S$^o$)5s $^5$S$_2^o$.  Bengston
and Miller estimate their errors to be 35\%.

%%Data for energy levels were taken from the tables of Basic
%%Atomic Spectroscopic Data found at the National Institute
%%for Standards and Technology website (NIST 2005).
%%THE NIST SITE HAS ALREADY BEEN CITED

%%For the three Br II lines,
%%the energy levels are attributed to the
%%unpublished work of Martin \& Tech (1984).
%%the center of gravity
%%wavelengths were computed from the energy levels and converted
%%from vacuum to air. formalism for these three lines.

%%We have approached the abundance analysis from two directions,
%%both using the model parameters T$_{eff}$ = 17500K, $\log(g) =
%%3.8$
%%and the abundance studies cited. In one approach, a LTE abundance
%%calculation
%%using ATLAS9/WIDTH9 (Kurucz 1993) was used to match the equivalent
%%widths.  The abundances are listed in the final column of Table 1,
%%and are in
%%good agreement with one another.
Results from ATLAS9/WIDTH9 are given in the final
column of Table 1.
They are based on the average of four
equivalent width measurements, two at Lund and two at Michigan.
A straight average of the logarithms is $-6.81 \pm 0.10$, using
a microturbulence of
$\xi_t = 1.0$ km s$^{-1}$.
The difference in
results using $\xi_t = 0$ and $\xi_t = 2$ ranges from 0.01
to 0.02 dex for the three lines.  We consider this
difference insignificant at the current level of accuracy.

In the spectral synthesis, hyperfine structure was 
incorporated for the wavelengths and oscillator
strengths using the magnetic hyperfine $A$ constants 
measured by Ranade (1951).
%%The second approach was synthetic spectrum fitting, in which
%%synthetic
Spectra
were generated using the SYNTHE program (Kurucz 1993). 
%%with the hfs components incorporated into the line list. 
Figure 1 presents these results for the case of
$v\cdot\sin(i) = 2.0$ km s$^{-1}$ and a bromine abundance of
$-6.8$, claimed
to be the best fit for the Br II lines. For each of the 
two segments (437B\_a, 437B\_b) of the observation the 
same two synthetic spectra are compared; accounting for 
the inclusion and exclusion of hfs. It can be seen that 
the hfs does broaden this Br II line at the expense of 
making the profile more shallow by an amount that is 
equivalent to about 0.1 dex. The seemingly good fit to the 
lower observation in the panel by the hfs excluded line 
profile is, of course, illusory, and it can be seen that 
the two spectra are noise affected.

The ``solar'' value, based on CI chondrites, is $-9.4$
(Lodders 2003, Asplund, Grevesse, and Sauval 2004).
The bromine excess in 3 Cen A  from both methods is thus
2.6 dex, only slightly less than the excesses found by 
Castelli, Parthasary, \& Hack for Ga and Kr. This 
abundance enhancement is computed under the assumption 
that both isotopes are present in the stellar atmosphere. 

Elemental abundance and isotopic anomalies are common for
chemically peculiar stars of the upper main sequence. For 
3 Cen A, isotope anomalies are inferred from line profiles 
for the elements He (Sargent and Jugaku 1961)
and Hg (Wahlgren and Hubrig 2004),
and may well exist for other
elements. In the event that the isotopic composition of Br 
in the atmosphere is that of only one isotope, the 
abundance enhancement--for that isotope--would be roughly
doubled.

\begin{figure}
%% \vspace{7.5cm}
%%  \resizebox{\hsize}{!}{\includegraphics{BrII4704.ps}}
  \resizebox{\hsize}{!}{\includegraphics{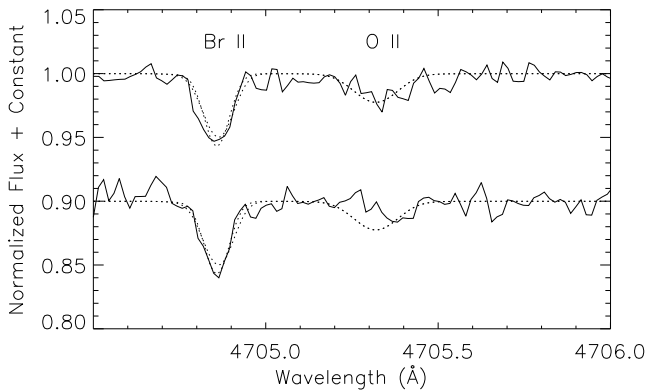}}
  \caption{Br II in the spectrum of 3 Cen A. The observation (solid)
is compared with synthetic spectrum calculations (dashed). Upper and
lower comparisons are for the two spectra of the observation. See
text for details.
}
  \label{BrII11}
\end{figure}

\section{Vanadium}

The strongest V II lines lie short of the traditional photographic
region (roughly $\lambda\lambda$3750-4650).
Table 2 shows our results for 31 V II lines.  The lines are primarily
strong transitions from Meggers, Corliss, \& Scribner (1975).
Wavelengths and transition probabilities are from the VALD
database (Ryabchikova, et al. 1999).  Details for individual
lines may be found at the VALD website.  The primary
reference for oscillator strengths is Biemont, {\it et al.}
(1989).

Many of the equivalent widths were measured
twice at Lund and twice at Michigan.  The number of
measurements is shown in the column headed `No.'  Those lines
with fewer than four measurements reflect a variety of circumstances
related to blending and the quality of the data used at Michigan
and at Lund.  Stellar wavelengths are given in the column headed
$\lambda^*$, with only the fraction of an angstrom shown.  We
consider $\pm 0.02$ \AA\ to be an acceptable deviation of the stellar
and laboratory position.  Larger deviations are flagged with an
asterisk; they usually indicate blending, or a setting error.
All equivalent widths were measured with respect to the local
continuum; those marked with a minus sign have been augmented
slightly to compensate for lowering of the local continuum by blends.
For example,
the largest such change was for $\lambda$3847, where the
equivalent widths was increased by 7\% because the
continuum was depressed by 7\% by the wing of H9.

\begin{figure}
%  \vspace{7.5cm}
%%  \resizebox{\hsize}{!}{\includegraphics{VII3102.ps}}
  \resizebox{\hsize}{!}{\includegraphics{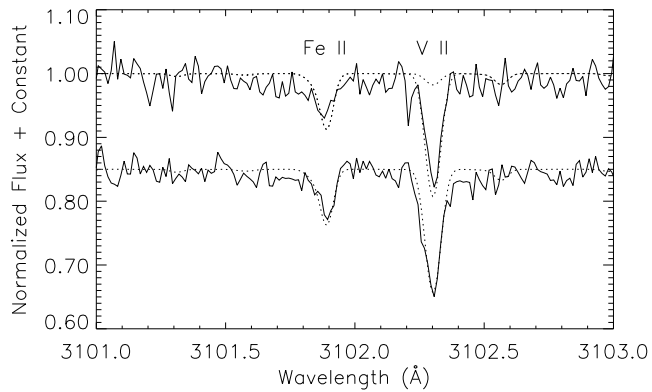}}
  \caption{V II in the spectrum of 3 Cen A. The observation (solid)
is compared with the synthetic spectrum calculation for best fit
abundance (upper and lower) and the solar abundance of
vanadium (upper only).
}
  \label{VII11}
\end{figure}

\begin{table}
\caption[]{V II lines studied}
\label{tab:v2}
\begin{tabular}{llrcrrr}
\hline\hline\\[-4pt]
\multicolumn{1}{c}{$\lambda$} &\multicolumn{1}{c}{$\lambda^*$}
&\multicolumn{1}{c}{$W_\lambda$} &\multicolumn{1}{c}{No.}
&\multicolumn{1}{c}{log$gf$} &\multicolumn{2}{c}{log(Ab/Sum)}
\\
\multicolumn{1}{c}{\AA\ }    &\multicolumn{1}{c}{\AA\ }
&\multicolumn{1}{c}{m\AA\ }  &\multicolumn{1}{c}{}
&\multicolumn{1}{c}{} &\multicolumn{1}{c}{$\xi_t$=0}
&\multicolumn{1}{c}{$\xi_t$=2}\\
\hline\\[-4pt]
3093.105 & 0.09 &  21.2 &   4  &  0.599 &   -6.57 &  -6.72  \\
3094.216 & 0.20$^a$&2.8 &   4  & -0.768 &   -6.93 &  -6.95  \\
3102.289 & 0.30 &  14.4 &   4  &  0.448 &   -6.77 &  -6.85  \\
3110.700 & 0.71 &   7.6 &   4  &  0.340 &   -7.07 &  -7.10  \\
3118.373 & 0.39 &   7.6 &   4  &  0.196 &   -6.93 &  -6.97  \\
3121.136 & 0.16 &   2.9 &   3  & -0.429 &   -6.78 &  -6.79  \\
3122.908 & 0.90 &   5.2 &   2  &  0.160 &   -6.16 &  -6.18  \\
3125.276 & 0.27$^b$&13.6&   4  &  0.044 &   -6.42 &  -6.49  \\
3126.211 & 0.19 &   5.6 &   4  & -0.268 &   -6.62 &  -6.64  \\
3130.261 & 0.25 &   3.8 &   3  & -0.285 &   -6.81 &  -6.82  \\
3133.327 & 0.31 &   3.5 &   3  & -0.484 &   -6.65 &  -6.67  \\
3134.933 & 0.94 &   3.7 &   4  &  0.050 &   -6.35 &  -6.37  \\
3188.514 & 0.51 &   4.8 &   4  &  0.089 &   -6.78 &  -6.80  \\
3190.683 & 0.70 &  11.4 &   2  &  0.262 &   -6.45 &  -6.51  \\
3217.113 & 0.09 &   4.1 &   4  &  0.198 &   -6.69 &  -6.71  \\
3254.762 & 0.77 &   2.6 &   3  & -0.042 &   -6.68 &  -6.69  \\
3267.704 & 0.70 &   6.1 &   4  &  0.283 &   -6.84 &  -6.87  \\
3271.123 & 0.11$^c$&6.5 &   2  &  0.377 &   -6.89 &  -6.92  \\
3517.296 & 0.34*$^d$&4.5&   2  & -0.208 &   -6.44 &  -6.46  \\
3545.194 & 0.19 &   1.9 &   2  & -0.259 &   -6.82 &  -6.82  \\
3556.792 & 0.81 &   2.4 &   4  & -0.066 &   -6.89 &  -6.90  \\
3589.749 & 0.70*&   1.8 &   2  & -0.295 &   -6.81 &  -6.81  \\
3715.466 & 0.44$^e$&1.9-&   2  & -0.250 &   -7.20 &  -7.20  \\
3727.343 & 0.33$^c$&2.5 &   3  & -0.231 &   -7.05 &  -7.06  \\
3815.388 & 0.36 &   1.9 &   2  & -0.400 &   -6.58 &  -6.58  \\
3847.339 & 0.38*&   1.5-&   4  & -0.608 &   -6.53 &  -6.53  \\
3878.704 & 0.72 &   3.5-&   3  & -0.609 &   -6.47 &  -6.48  \\
3899.129 & 0.12 &   2.5-&   2  & -0.784 &   -6.46 &  -6.46  \\
3903.262 & 0.29 &   1.9-&   2  & -0.938 &   -6.54 &  -6.55  \\
4005.705 & 0.70 &   2.9 &   4  & -0.522 &   -6.64 &  -6.65  \\
4023.378 & 0.40 &   2.1-&   2  & -0.689 &   -6.63 &  -6.63  \\
         &      &       &      &        &         &         \\
%%         &      &       &      & Ave =  &   -6.69 &  -6.72  \\
%%         &      &       &      & $\sigma$ =  &0.23&   0.23  \\
%%\\
\hline\\[-4pt]
\end{tabular}
\begin{list}{}{}
\item{Notes:  $a$, asymmetric; $b$, broad; $c$, poor continuum;
$d$, blend; $e$, Balmer wing; *, wavelength discrepancy; -, equivalent width
augmented (see text)}
\end{list}
\end{table}

Abundances based on the equivalent width measurements are presented
for two values of the microturbulent velocity ($\xi_t$). Averages
of the logarithms yield
$\rm \log(V/Sum) = -6.7 \pm 0.2$ for the
two microturbulences.  A subset the 5 lines from the principal
diagonal of V II Multiplet 1 yields an abundance of $-6.8$,
about 0.1 dex lower.  These are the lines selected for spectral
synthesis at Lund.

%%If we select only those lines with four
%%equivalent width measurements, and exclude those with wavelength
%%mismatches or profile asymmetries, we obtain averages of 6.74 and

%%The vanadium abundance was determined
%%at Lund from line profile fitting,
%%as for bromine.

Terrestrial vanadium has two stable
isotopes in
the proportions $^{50}$V:$^{51}$V = 0.250:99.750
(Rosman and Taylor 1998). The even-A
isotope has been ignored in our calculations. The odd-A isotope
possesses hfs for its line profiles. The hfs $A$ constants in V II
have been
determined by Arvidsson (2003) for 26 levels from laboratory Fourier
transform spectroscopy experiments, with hfs components and their
relative intensities tabulated.

Five lines of V II in our stellar
spectrum ($\lambda\lambda$3093.105, 3102.289, 3110.700, 3118.373,
3125.276) have available hfs data, and could be included into our
synthetic spectrum calculations. All five lines are well fit with an
abundance of $-6.84 \pm 0.05$.

%%IT DOESN'T SEEM LIKELY THAT THIS AGREEMENT IS BY CHANCE.
%%WHAT SEEMS PUZZLING TO ME IS HOW THE SCATTER ARISES

%%For these same five lines
%%the results from the equivalent width analysis yield
%%-6.75 and -6.83 for $\xi_t = 0.0$ and 2.0 km/s respectively.
%%the same average abundance ($-$6.83) but with a

The scatter from the equivalent widths is much larger
than from the synthesis, mostly resulting
from noise and line blending. The hfs does not make a noticeable
effect upon these particular lines.
%%We adopt $-6.84$ from
%%the syntheses.

Figure 2 presents a comparison of the
observed data with synthetic spectra for the line V II $\lambda$3102
for the two spectra of observation segment 346B.The upper comparison
in the figure also includes the synthetic spectrum computed with the
solar abundance of vanadium.

We adopt a vanadium abundance of $-6.84$, giving a slight
preference to the profile fitting of the stronger lines
of Multiplet 1.  This
%%The vanadium abundance determined
is an excess of 1.2 dex above the
corresponding solar value ($-8.0$).
It is similar to the excesses reported
by Adelman \& Pintado (2000) or Castelli, Parthasarathy, \& Hack
(1997) for the odd-Z elements P, Sc, and Mn in 3 Cen A.

\section{Conclusions}

The numerous odd-Z anomalies of 3 Cen A, along with the
dominance of $\rm ^3He$, make this object a paradigm
example of chemical separation in stars.
A comprehensive study
that would simultaneously account for all of these
peculiarities would have great value.

\begin{acknowledgements}
  CRC gratefully acknowledges help from Dr. R. L. Kurucz with the
WIDTH code.  Thanks are also due to the ESO staff for the UVESPOP
public data archive.
\end{acknowledgements}
%%\begin{samepage}

\end{document}